\documentclass[prb,superscriptaddress,showpacs,floatfix,tightenlines,10pt,twocolumn]{revtex4}
\usepackage{amssymb}
\usepackage{amsmath}
\usepackage{graphicx}

\renewcommand{\Re}{\operatorname{Re}}
\renewcommand{\Im}{\operatorname{Im}}
\begin{document}

\title{Charge density wave with meronlike spin texture induced by a lateral
superlattice in a two-dimensional electron gas}
\author{R. C\^{o}t\'{e}}
\affiliation{D\'{e}partement de physique, Universit\'{e} de Sherbrooke, Sherbrooke, Qu%
\'{e}bec, J1K 2R1, Canada}
\author{Xavier Bazier-Matte}
\affiliation{D\'{e}partement de physique, Universit\'{e} de Sherbrooke, Sherbrooke, Qu%
\'{e}bec, J1K 2R1, Canada}
\keywords{}
\pacs{73.22Gk,73.43.Lp,73.43.-f}

\begin{abstract}
The combined effect of a lateral square superlattice potential and the
Coulomb interaction on the ground state of a two-dimensional electron gas in
a perpendicular magnetic field is studied for different rational values of $%
\Gamma ,$ the inverse of the number of flux quanta per unit cell of the
external potential, at filling factor $\nu =1$ in Landau level $N=0.$ When
Landau level mixing and disorder effects are neglected, increasing the
strength $W_{0}$ of the potential induces a transition, at a critical
strength $W_{0}^{\left( c\right) },$ from a uniform and fully spin polarized
state to a two-dimensional charge density wave (CDW) with a meronlike spin
texture at each maximum and minimum of the CDW. The collective excitations
of this \textquotedblleft vortex-CDW\textquotedblright\ are similar to those
of the Skyrme crystal that is expected to be the ground state \textit{near}
filling factor $\nu =1$. In particular, a broken U(1) symmetry in the
vortex-CDW results in an extra gapless phase mode that could provide a fast
channel for the relaxation of nuclear spins. The average spin polarization $%
S_{z}$ changes in a continuous or discontinuous manner as $W_{0}$ is
increased depending on whether $\Gamma \in \left[ 1/2,1\right] $ or $\Gamma
\in \left[ 0,1/2\right] .$ The phase mode and the meronlike spin texture
disappear at large value of $W_{0},$ leaving as the ground state partially
spin-polarized CDW if $\Gamma \neq 1/2$ or a spin-unpolarized CDW if $\Gamma
=1/2.$
\end{abstract}

\date{\today }
\maketitle

\section{INTRODUCTION}

The two-dimensional electron gas (2DEG)\ in a perpendicular magnetic field
has a very rich phase diagram that includes several phases such as the
Laughlin liquids that give rise to the integer and quantum Hall effects\cite%
{Girvin1}, the Wigner crystal at small filling factor in each Landau level%
\cite{Wigner,Chui}, the bubble crystals and the stripe phase in higher
Landau levels\cite{Fogler} and the Skyrme crystal\cite{Sondhi,SkyrmeCrystal}
near filling factor $\nu =1$ in the lowest Landau level. The phase diagram
is even more complex when system with extra degrees of freedom such as
double quantum wells (DQWs) are considered\cite{Pinkzuk}. In DQWs, the
orientation of the pseudospin vector associated with the layer degree of
freedom can be modified by changing the tunneling and electrical bias
between the layers.

Another way to modify the properties of the 2DEG is by the addition of a
lateral two-dimensional superlattice patterned on top of the GaAs/AlGaAs
heterojunction hosting the 2DEG that creates a spatially modulated potential
at the position of the 2DEG\cite{ReviewModulation}. The effect of a
one-dimensional periodic potential on the Landau levels is particularly
interesting\cite{Rauh,Gerhardts} since it leads to commensurability problems
due to the presence of different lengths scales: the lattice constant of the
external potential $a_{0},$ the magnetic length $\ell =\sqrt{\hslash c/eB}$ (%
$B$ is the magnetic field) and the Fermi wavelength. Novel magnetoresistance
oscillations with period different than that of the well-known Shubnikov-de
Haas oscillations have been detected in such systems. Even more interesting
is the effect of a periodic two-dimensional potential on the band structure
of the 2DEG\cite{G2,Thouless,Gumbs}. The intricate pattern of eigenvalues
that results from such potential has been studied by many authors and is
known as the Hofstadter butterfly spectrum\cite{Hofstadter}. Its observation
in GaAs/AlGaAs heterojunction is very difficult due to screening and
disorder effects but experimental signature in magnetotransport experiments
in 2DEGs with a lateral surface superlattice potential with period of the
order of $100$ nm and less have been reported\cite{Albrecht, Melinte,Experi}%
. Interest in this problem has been revived recently by the experimental
observations of the Hofstadter's butterfly spectrum that use the moir\'{e}
superlattices that arise from graphene or bilayer graphene placed on top of
hexagonal boron nitride\cite{Dean,Hunt}. Another interest of superlattice
potentials is their use to create artificial lattices. For example, a
lateral superlattice with a honeycomb crystal structure has recently been
proposed to create an artificial graphenelike system\cite{Gibertini, Park}
in a GaAs/AlGaAs heterojunction.

In this work, we study theoretically the effect of a square lattice lateral
potential with a period $a_{0}$ on the ground state of the 2DEG in
GaAs/AlGaAs heterojunction at filling factor $\nu =1$ and in Landau level $%
N=0.$ We include the spin degree of freedom and use the Hartree-Fock
approximation to study the combined effects of the external potential and
the Coulomb interaction. We assume that the potential is sufficiently weak
so that Landau level mixing can be neglected. We also ignore disorder and
work at zero temperature. We vary the potential strength $W_{0}$ and
calculate the ground state for different rational values of $\Gamma =\varphi
_{0}/Ba_{0}^{2}=q/p\in \left[ 0,1\right] $ (where $\varphi _{0}=hc/e$ is the
flux quantum and $q,p$ are integers with no common factors) which is the
inverse of the number of flux quanta per unit cell of the surface potential.
Our formalism allows for the formation of uniform as well as spatially
modulated ground states with or without spin texture. Our calculation
indicates that, at a critical value, $W_{0}^{\left( c\right) },$ of the
external potential, there is a transition from a uniform fully spin
polarized state to a charge density wave (CDW) with an intricate spin
texture. Each unit cell of this CDW contains two positive and two negative
amplitude modulations and the vortex spin texture at each maximum(minimum)\
resembles that of a positively(negatively) charged meron. The two
positively(negatively) charged merons in each unit cell have the same
vorticity but a global phase that differs by $\pi .$ These meronlike
textures, however, are not quantized since the amplitude of the CDW varies
continuously with $W_{0}$. In the vortex-CDW, as we call it, the average
spin polarization $S_{z}$ varies with $W_{0}$ in a continuous or
discontinuous manner depending on whether $\Gamma \in \left[ 1/2,1\right] $
or $\Gamma \in \left[ 0,1/2\right] $ and saturates at a finite, positive,
value of $S_{z}$ that depends only on $\Gamma $ in most cases. In the
special case $\Gamma =1/2,$ the vortex-CDW phase is absent and the
transition is directly from a fully spin polarized and uniform 2DEG to an
unpolarized CDW. The phase diagram for $\Gamma \in \left[ 0,1/2\right] $ is
richer than that of $\Gamma \in \left[ 1/2,1\right] $ as it involves the
transition between the vortex-CDW and its conjugate phase, the anti-vortex
CDW, obtained by reversing $S_{z}\left( \mathbf{r}\right) $ and inverting
the vorticity of all merons. This transition between the two CDWs is
accompanied by a discontinuous change of $S_{z}$ that becomes continuous
when the Zeeman coupling goes to zero.

We study the properties of the vortex-CDW at different values of $\Gamma $
and with a particular emphasis on its collective excitations which we derive
using the generalized random-phase approximation (GRPA). The vortex-CDW has
collective modes that have much in common with the collective excitations of
the Skyrme crystal\cite{CoteSkyrme} that is expected to be the ground state
near filling factor $\nu =1$ in $N=0.$ Namely, the broken U(1) symmetry in
the vortex-CDW phase leads to a new gapless mode that can provide a fast
channel for the relaxation of nuclear spins\cite{Cote3}. This mode and the
meronlike spin texture disappear at larger values of the external potential
leaving a ground state that is either unpolarized if $\Gamma =1/2$ or
partially polarized if $\Gamma \neq 1/2.$

Our paper is organized as follows. In Sec. II, we introduce the Hamiltonian
of the 2DEG in the presence of the lateral square lattice potential and
briefly review the Hartree-Fock and generalized random-phase approximation
that we use to compute the density of states, the density and spin profiles
and the collective excitations of the various phases. In Sec. III, we
present our numerical results for the phase diagram of the 2DEG as a
function of the potential strength $W_{0}$ and the inverse magnetic flux per
unit cell $\Gamma $. We conclude in Sec. IV\ with a discussion on the
experimental detection of the new vortex-CDW state.

\section{HAMILTONIAN OF THE 2DEG IN AN EXTERNAL POTENTIAL}

The system we consider is a 2DEG in a GaAs/AlGaAs heterojunction or quantum
well submitted to a perpendicular magnetic field $\mathbf{B}=B\widehat{%
\mathbf{z}}$ and to a lateral superlattice potential $V_{e}\left( \mathbf{r}%
\right) .$ The coupling of the electrons to this external potential is given
by $H_{e}=-e\int d\mathbf{r}V_{e}\left( \mathbf{r}\right) n_{e}\left( 
\mathbf{r}\right) $, where $n_{e}\left( \mathbf{r}\right) =\sum_{\alpha =\pm
}n_{e,\alpha }\left( \mathbf{r}\right) $ is the total density operator
including both spin states $\alpha =\pm 1$ (we take $e>0$)$.$ We assume that
only the Landau level $N=0$ is occupied but our calculation can easily be
generalized to any Landau level by changing the effective interactions $%
H\left( \mathbf{q}\right) $ and $X\left( \mathbf{q}\right) $ and the form
factor $F\left( \mathbf{q}\right) .$ The Hamiltonian of the interacting 2DEG
is given, in the Hartree-Fock approximation, by 
\begin{eqnarray}
H_{HF} &=&-N_{\varphi }\frac{\Delta _{Z}}{2}\sum_{\alpha }\alpha \rho
_{\alpha ,\alpha }\left( 0\right)  \label{hamiltonien} \\
&&-\frac{eN_{\varphi }}{S}\sum_{\mathbf{q}}\sum_{\alpha }V_{e}\left( -%
\mathbf{q}\right) F\left( -\mathbf{q}\right) \rho _{\alpha ,\alpha }\left( 
\mathbf{q}\right)  \notag \\
&&+N_{\varphi }\sum_{\alpha ,\beta }\sum_{\mathbf{q}\neq 0}H\left( \mathbf{q}%
\right) \left\langle \rho _{\alpha ,\alpha }\left( -\mathbf{q}\right)
\right\rangle \rho _{\beta ,\beta }\left( \mathbf{q}\right)  \notag \\
&&-N_{\varphi }\sum_{\alpha ,\beta }\sum_{\mathbf{q}}X\left( \mathbf{q}%
\right) \left\langle \rho _{\alpha ,\beta }\left( -\mathbf{q}\right)
\right\rangle \rho _{\beta ,\alpha }\left( \mathbf{q}\right) ,  \notag
\end{eqnarray}%
where $S$ is the 2DEG area, $N_{\varphi }=S/2\pi \ell ^{2}$ is the Landau
level degeneracy, and the form factor for the $N=0$ Landau level is%
\begin{equation}
F\left( \mathbf{q}\right) =e^{-q^{2}\ell ^{2}/2},  \label{form}
\end{equation}%
where $\ell =\sqrt{\hslash c/eB}$ is the magnetic length. The averages are
over the Hartree-Fock ground state of the 2DEG. The non-interacting
single-particle energies, measured with respect to the kinetic energy $%
\hslash \omega _{c}/2,$ are given by%
\begin{equation}
E_{\alpha }=\alpha \frac{\Delta _{Z}}{2},
\end{equation}%
where the Zeeman energy $\Delta _{Z}=\left\vert g^{\ast }\right\vert \mu
_{B}B,$ with $g^{\ast }$ the effective $g-$factor of bulk GaAs and $\mu _{B}$
the Bohr magneton. In some experiments on skyrmions, the effective $g-$%
factor was tuned in the range $-0.11$ to $0.065$ by applying hydrostatic
pressure to a sample of GaAs/AlGaAs modulation doped quantum well\cite%
{Potemski}. In our study, we will thus consider that $\Delta _{Z}$ is not
determined by the magnetic field, but is instead a parameter than can be
adjusted.

The Hartree and Fock interactions in $N=0$ are given by%
\begin{eqnarray}
H\left( \mathbf{q}\right) &=&\left( \frac{e^{2}}{\kappa \ell }\right) \frac{1%
}{q\ell }e^{-q^{2}\ell ^{2}/2},  \label{inter} \\
X\left( \mathbf{q}\right) &=&\left( \frac{e^{2}}{\kappa \ell }\right) \sqrt{2%
}\int_{0}^{\infty }dxe^{-x^{2}}J_{0}\left( \sqrt{2}xq\ell \right) ,  \notag
\end{eqnarray}%
where $\kappa =12.9$ is the dielectric constant of GaAs. Finally, the
operators $\rho _{\alpha ,\beta }\left( \mathbf{q}\right) $ are defined by%
\begin{eqnarray}
\rho _{\alpha ,\beta }\left( \mathbf{q}\right) &\equiv &\frac{1}{N_{\varphi }%
}\sum_{X,X^{\prime }}e^{-\frac{i}{2}q_{x}\left( X+X^{\prime }\right) }
\label{rho} \\
&&\times \delta _{X,X^{\prime }+q_{y}\ell ^{2}}c_{X,\alpha }^{\dagger
}c_{X^{\prime },\beta },  \notag
\end{eqnarray}%
where $c_{X,\alpha }^{\dagger }$ is the operator that creates an electron
with guiding-center index $X$ (in the Landau gauge) and spin $\alpha .$ The
four operators $\rho _{\alpha ,\beta }\left( \mathbf{q}\right) $ are related
to the averaged electronic and spin densities in the $xy-$plane by%
\begin{eqnarray}
n_{\alpha }\left( \mathbf{r}\right) &=&\frac{1}{2\pi \ell ^{2}}\sum_{\mathbf{%
q}}\left\langle \rho _{\alpha ,\alpha }\left( \mathbf{q}\right)
\right\rangle e^{-q^{2}\ell ^{2}/4}e^{i\mathbf{q}\cdot \mathbf{r}}, \\
S_{x}\left( \mathbf{r}\right) &=&\frac{1}{2\pi \ell ^{2}}\sum_{\mathbf{q}}%
\Re\left[ \left\langle \rho _{+,-}\left( \mathbf{q}\right)
\right\rangle e^{-q^{2}\ell ^{2}/4}e^{i\mathbf{q}\cdot \mathbf{r}}\right] ,
\\
S_{y}\left( \mathbf{r}\right) &=&\frac{1}{2\pi \ell ^{2}}\sum_{\mathbf{q}}%
\Im\left[ \left\langle \rho _{+,-}\left( \mathbf{q}\right)
\right\rangle e^{-q^{2}\ell ^{2}/4}e^{i\mathbf{q}\cdot \mathbf{r}}\right] ,
\\
S_{z}\left( \mathbf{r}\right) &=&\frac{\hslash }{2}\left[ n_{+}\left( 
\mathbf{r}\right) -n_{-}\left( \mathbf{r}\right) \right] .
\end{eqnarray}%
The $\left\langle \rho _{\alpha ,\beta }\left( \mathbf{q}\right)
\right\rangle ^{\prime }s$ can be considered as the order parameters of an
ordered phase of the 2DEG.

The averaged Hartree-Fock ground-state energy per electron at $\nu =1$ is
given by%
\begin{eqnarray}
\frac{\left\langle H_{HF}\right\rangle }{N_{e}} &=&-\frac{\Delta _{Z}}{2}%
\sum_{\alpha }\alpha \left\langle \rho _{\alpha ,\alpha }\left( 0\right)
\right\rangle  \label{energy} \\
&&-\frac{1}{S}\sum_{\mathbf{q}}\sum_{\alpha }V_{e}\left( -\mathbf{q}\right)
F\left( \mathbf{q}\right) \left\langle \rho _{\alpha ,\alpha }\left( \mathbf{%
q}\right) \right\rangle  \notag \\
&&+\frac{1}{2}\sum_{\alpha ,\beta }\sum_{\mathbf{q}\neq 0}H\left( \mathbf{q}%
\right) \left\langle \rho _{\alpha ,\alpha }\left( -\mathbf{q}\right)
\right\rangle \left\langle \rho _{\beta ,\beta }\left( \mathbf{q}\right)
\right\rangle  \notag \\
&&-\frac{1}{2}\sum_{\alpha ,\beta }\sum_{\mathbf{q}}X\left( \mathbf{q}%
\right) \left\vert \left\langle \rho _{\alpha ,\beta }\left( \mathbf{q}%
\right) \right\rangle \right\vert ^{2}.  \notag
\end{eqnarray}

The order parameters $\left\langle \rho _{\alpha ,\beta }\left( \mathbf{q}%
\right) \right\rangle $ are computed by solving the Hartree-Fock equation
for the single-particle Green's function $G_{\alpha ,\beta }\left( \mathbf{q,%
}\tau \right) $ which is defined by%
\begin{eqnarray}
G_{\alpha ,\beta }\left( \mathbf{q,}\tau \right) &=&\frac{1}{N_{\varphi }}%
\sum_{X,X^{\prime }}e^{-\frac{i}{2}q_{x}\left( X+X^{\prime }\right) } \\
&&\times \delta _{X,X^{\prime }-q_{y}\ell ^{2}}G_{\alpha ,\beta }\left(
X,X^{\prime },\tau \right) ,  \notag
\end{eqnarray}%
where%
\begin{equation}
G_{\alpha ,\beta }\left( X,X^{\prime },\tau \right) =-\left\langle
Tc_{X,\alpha }\left( \tau \right) c_{X^{\prime },\beta }^{\dagger }\left(
0\right) \right\rangle .
\end{equation}%
They are obtained with the relation 
\begin{equation}
\left\langle \rho _{\alpha ,\beta }\left( \mathbf{q}\right) \right\rangle
=G_{\beta ,\alpha }\left( \mathbf{q,}\tau =0^{-}\right) .
\end{equation}

The Hartree-Fock equation of motion for the Green's function $G_{\alpha
,\beta }\left( \mathbf{q},i\omega _{n}\right) $ is given\cite{CoteMethode}
by 
\begin{eqnarray}
&&\left[ i\omega _{n}-\frac{1}{\hslash }\left( \alpha \frac{\Delta _{Z}}{2}%
-\mu \right) \right] G_{\alpha ,\beta }\left( \mathbf{q},i\omega _{n}\right)
\label{motion} \\
&=&\delta _{\mathbf{q},0}\delta _{\alpha ,\beta }  \notag \\
&&-\frac{e}{\hslash S}\sum_{\mathbf{q}^{\prime }}V_{e}\left( \mathbf{q}-%
\mathbf{q}^{\prime }\right) F\left( \left\vert \mathbf{q}-\mathbf{q}^{\prime
}\right\vert \right)  \notag \\
&&\times \gamma _{\mathbf{q},\mathbf{q}^{\prime }}G_{\alpha ,\beta }\left( 
\mathbf{q}^{\prime },i\omega _{n}\right)  \notag \\
&&+\frac{1}{\hslash }\sum_{\mathbf{q}^{\prime }\neq \mathbf{q}}U^{H}\left( 
\mathbf{q-q}^{\prime }\right) \gamma _{\mathbf{q},\mathbf{q}^{\prime
}}G_{\alpha ,\beta }\left( \mathbf{q}^{\prime },i\omega _{n}\right)  \notag
\\
&&-\frac{1}{\hslash }\sum_{\mathbf{q}^{\prime }}\sum_{\gamma }U_{\alpha
,\gamma }^{F}\left( \mathbf{q-q}^{\prime }\right) \gamma _{\mathbf{q},%
\mathbf{q}^{\prime }}G_{\gamma ,\beta }\left( \mathbf{q}^{\prime },i\omega
_{n}\right) ,  \notag
\end{eqnarray}%
where%
\begin{equation}
\gamma _{\mathbf{q},\mathbf{q}^{\prime }}=e^{-i\left( \mathbf{q}\times 
\mathbf{q}^{\prime }\right) \cdot \widehat{\mathbf{z}}\ell ^{2}/2},
\end{equation}%
and $\omega _{n}$ are fermionic Matsubara frequencies, $\mu $ is the
chemical potential and we have defined the potentials

\begin{eqnarray}
U^{H}\left( \mathbf{q}\right) &=&\sum_{\alpha }H\left( \mathbf{q}\right)
\left\langle \rho _{\alpha ,\alpha }\left( \mathbf{q}\right) \right\rangle ,
\\
U_{\alpha ,\beta }^{F}\left( \mathbf{q}\right) &=&X\left( \mathbf{q}\right)
\left\langle \rho _{\beta ,\alpha }\left( \mathbf{q}\right) \right\rangle .
\end{eqnarray}

These potentials depend on the order parameters $\left\langle \rho _{\alpha
,\beta }\left( \mathbf{q}\right) \right\rangle $ that are unknown. The
equation of motion for $G_{\alpha ,\beta }\left( \mathbf{q},i\omega
_{n}\right) $ must thus be solved numerically\cite{CoteMethode} by using a
seed for the order parameters and then iterate Eq. (\ref{motion}) until a
convergent solution is found. In case several solutions are found
(corresponding to different choice for the initial seed), we choose the one
with the lowest energy and compute the dispersion relation of its collective
modes to make sure that it is a stable solution. We remark that this method
does not guarantee that the true ground state is the solution that we keep.

The density of states $g\left( \omega \right) $ is obtained from the
single-particle Green's function by using the relation 
\begin{equation}
g\left( \omega \right) =-\frac{N_{\varphi }}{\pi }\sum_{\alpha }\Im%
\left[ G_{\alpha ,\alpha }\left( \mathbf{q}=0,\omega +i\delta \right) \right]
.
\end{equation}

To find the dispersion relation of the collective modes, we derive the
equation of motion in the generalized random-phase approximation for the
two-particle Green's function%
\begin{eqnarray}
\chi _{\alpha ,\beta ,\gamma ,\delta }\left( \mathbf{q,q}^{\prime };\tau
\right) &=&-N_{\varphi }\left\langle T\rho _{\alpha ,\beta }\left( \mathbf{q,%
}\tau \right) \rho _{\gamma ,\delta }\left( -\mathbf{q}^{\prime },0\right)
\right\rangle \\
&&+N_{\varphi }\left\langle \rho _{\alpha ,\beta }\left( \mathbf{q}\right)
\right\rangle \left\langle \rho _{\gamma ,\delta }\left( -\mathbf{q}^{\prime
}\right) \right\rangle .  \notag
\end{eqnarray}%
This equation is%
\begin{eqnarray}
&&\chi _{\alpha ,\beta ,\gamma ,\delta }\left( \mathbf{q},\mathbf{q}^{\prime
};i\Omega _{n}\right)  \label{GRPA} \\
&=&\chi _{\alpha ,\beta ,\gamma ,\delta }^{\left( 0\right) }\left( \mathbf{q}%
,\mathbf{q}^{\prime };i\Omega _{n}\right)  \notag \\
&&+\frac{1}{\hslash }\sum_{\xi ,\lambda }\sum_{\mathbf{q}^{\prime \prime
}}\chi _{\alpha ,\beta ,\xi ,\xi }^{\left( 0\right) }\left( \mathbf{q},%
\mathbf{q}^{\prime \prime };i\Omega _{n}\right) H\left( \mathbf{q}^{\prime
\prime }\right) \chi _{\lambda ,\lambda ,\gamma ,\delta }\left( \mathbf{q}%
^{\prime \prime },\mathbf{q}^{\prime };i\Omega _{n}\right)  \notag \\
&&-\frac{1}{\hslash }\sum_{\xi ,\lambda }\sum_{\mathbf{q}^{\prime \prime
}}\chi _{\alpha ,\beta ,\xi ,\lambda }^{\left( 0\right) }\left( \mathbf{q},%
\mathbf{q}^{\prime \prime };i\Omega _{n}\right) X\left( \mathbf{q}^{\prime
\prime }\right) \chi _{\lambda ,\xi ,\gamma ,\delta }\left( \mathbf{q}%
^{\prime \prime },\mathbf{q}^{\prime };i\Omega _{n}\right) ,  \notag
\end{eqnarray}%
where $\Omega _{n}$ is a bosonic Matsubara frequency. Equation (\ref{GRPA})
represents the summation of bubble (polarization effects) and ladder
(excitonic corrections) diagrams. The Hartree-Fock two-particle Green's
function (the single-bubble Feynman diagram with Hartree-Fock propagators)
that enters this equation is obtained from the Hartree-Fock equation of
motion for $\chi _{\alpha ,\beta ,\gamma ,\delta }\left( \mathbf{q,q}%
^{\prime };\tau \right) $ and is given by

\begin{eqnarray}
&&\left[ i\hslash \Omega _{n}+\left( E_{\alpha }-E_{\beta }\right) \right]
\chi _{\alpha ,\beta ,\gamma ,\delta }^{\left( 0\right) }\left( \mathbf{q},%
\mathbf{q}^{\prime },i\Omega _{n}\right) \\
&=&\hslash \left[ \gamma _{\mathbf{q},\mathbf{q}^{\prime }}^{\ast
}\left\langle \rho _{\alpha ,\delta }\left( \mathbf{q-q}^{\prime }\right)
\right\rangle \delta _{\beta ,\gamma }-\gamma _{\mathbf{q},\mathbf{q}%
^{\prime }}\left\langle \rho _{\gamma ,\beta }\left( \mathbf{q-q}^{\prime
}\right) \right\rangle \delta _{\alpha ,\delta }\right]  \notag \\
&&-\frac{e}{S}\sum_{\mathbf{q}^{\prime \prime }}V_{e}\left( \mathbf{q}-%
\mathbf{q}^{\prime \prime }\right) F\left( \left\vert \mathbf{q}-\mathbf{q}%
^{\prime \prime }\right\vert \right) \left[ \gamma _{\mathbf{q},\mathbf{q}%
^{\prime \prime }}^{\ast }-\gamma _{\mathbf{q},\mathbf{q}^{\prime \prime }}%
\right]  \notag \\
&&\times \chi _{\alpha ,\beta ,\gamma ,\delta }^{\left( 0\right) }\left( 
\mathbf{q}^{\prime \prime },\mathbf{q}^{\prime },i\Omega _{n}\right)  \notag
\\
&&-\sum_{\mathbf{q}^{\prime \prime }\neq 0}U^{H}\left( \mathbf{q-q}^{\prime
\prime }\right) \left[ \gamma _{\mathbf{q},\mathbf{q}^{\prime \prime
}}^{\ast }-\gamma _{\mathbf{q},\mathbf{q}^{\prime \prime }}\right]  \notag \\
&&\times \chi _{\alpha ,\beta ,\gamma ,\delta }^{\left( 0\right) }\left( 
\mathbf{q}^{\prime \prime },\mathbf{q}^{\prime },i\Omega _{n}\right)  \notag
\\
&&+\sum_{\alpha ^{\prime }}\sum_{\mathbf{q}^{\prime \prime }}U_{\alpha
,\alpha ^{\prime }}^{F}\left( \mathbf{q-q}^{\prime \prime }\right) \gamma _{%
\mathbf{q},\mathbf{q}^{\prime \prime }}^{\ast }\chi _{\alpha ^{\prime
},\beta ,\gamma ,\delta }^{\left( 0\right) }\left( \mathbf{q}^{\prime \prime
},\mathbf{q}^{\prime },i\Omega _{n}\right)  \notag \\
&&-\sum_{\beta ^{\prime }}\sum_{\mathbf{q}^{\prime \prime }}U_{\beta
^{\prime },\beta }^{F}\left( \mathbf{q-q}^{\prime \prime }\right) \gamma _{%
\mathbf{q},\mathbf{q}^{\prime \prime }}\chi _{\alpha ,\beta ^{\prime
},\gamma ,\delta }^{\left( 0\right) }\left( \mathbf{q}^{\prime \prime },%
\mathbf{q}^{\prime },i\Omega _{n}\right) .  \notag
\end{eqnarray}%
By defining the super-indices $I=\left( \alpha ,\beta \right) $ and $%
J=\left( \gamma ,\delta \right) ,$ Eq. (\ref{GRPA}) can be rewritten as a $%
4\times 4$ matrix equation for the matrix of Green's functions $\chi _{I,J}.$
This equation has the form $\left[ i\Omega _{n}I-F\right] \chi =B.$ The
matrix $F$, which depends only on the $\left\langle \rho _{\alpha ,\beta
}\left( \mathbf{q}\right) \right\rangle ^{\prime }s$ is then diagonalized
numerically to find $\chi _{I,J}$. The retarded response functions are
obtained with the analytic continuation $i\Omega _{n}\rightarrow \omega
+i\delta .$ We compute the following density and spin responses:%
\begin{equation}
\chi _{\rho _{i},\rho _{j}}^{R}\left( \mathbf{q,q}^{\prime };\omega \right)
=-iN_{\phi }\left[ \left\langle \left[ \rho _{i}\left( \mathbf{q,}t\right)
,\rho _{j}\left( -\mathbf{q}^{\prime },t^{\prime }\right) \right]
\right\rangle \theta \left( t-t^{\prime }\right) \right] _{\omega },
\label{response}
\end{equation}%
where $i=n,x,y,z$ and the operators 
\begin{eqnarray}
\rho _{x} &=&\frac{1}{2}\left[ \rho _{+,-}+\rho _{-,+}\right] , \\
\rho _{x} &=&\frac{1}{2i}\left[ \rho _{+,-}-\rho _{-,+}\right] , \\
\rho _{z} &=&\frac{1}{2}\left[ \rho _{+,+}-\rho _{-,-}\right] , \\
\rho _{n} &=&\rho _{+,+}+\rho _{-,-}.
\end{eqnarray}

In a uniform phase, the order parameters $\left\langle \rho _{\alpha ,\beta
}\left( \mathbf{q}\right) \right\rangle $ are finite only when $\mathbf{q}=0$
while in a two-dimensional CDW, they can be non zero each time $\mathbf{q}=%
\mathbf{G}$, where $\mathbf{G}$ is a reciprocal lattice vector of the CDW.
For the response function, we have to compute $\chi _{\rho _{i},\rho
_{j}}^{R}\left( \mathbf{q,q};\omega \right) $ in the uniform phase and $\chi
_{\rho _{i},\rho _{j}}^{R}\left( \mathbf{k}+\mathbf{G,k+G}^{\prime };\omega
\right) $ in the CDW where $\mathbf{k}$ is, by definition, a vector in the
first Brillouin zone of the CDW. In the CDW, the GRPA\ matrices $F$ and $B$
have dimensions $4n_{R}\times 4n_{R}$, where $n_{R}$ is the number of
reciprocal lattice vectors considered in the numerical calculation. We
typically take $n_{R}\approx 600.$

The formalism developed in this section can also be applied to the 2DEG\ in
graphene if the electrons are assumed to occupy only one of the two valleys.
In Landau level $N=0$ of graphene (and only in this level), the form factor $%
F\left( \mathbf{q}\right) $ and the Hartree and Fock interactions $H\left( 
\mathbf{q}\right) $ and $X\left( \mathbf{q}\right) $ are the same as those
given in Eqs. (\ref{form},\ref{inter}).

\section{PHASE\ DIAGRAM FOR $\protect\nu =1$}

We study the phase diagram of the 2DEG at filling factor $\nu =1$ in Landau
level $N=0$ and at temperature $T=0$ K. For the external potential, we
choose the simple square lattice form%
\begin{equation}
V_{e}\left( \mathbf{r}\right) =2V_{e}\left[ \cos \left( \frac{2\pi }{\sqrt{2}%
a_{0}}\left( x+y\right) \right) +\cos \left( \frac{2\pi }{\sqrt{2}a_{0}}%
\left( x-y\right) \right) \right] ,
\end{equation}%
so that $V_{e}\left( \mathbf{q}\right) =SV_{e}\delta _{\mathbf{q},\mathbf{G}%
_{1}}$ in Eq. (\ref{hamiltonien}) with the vectors $\mathbf{G}_{1}\mathbf{%
\in }\frac{2\pi }{\sqrt{2}a_{0}}\left\{ \left( 1,1\right) ,\left(
1,-1\right) ,\left( -1,1\right) ,\left( -1,-1\right) \right\} .$ This
external potential tries to impose a two-dimensional density modulation of
the 2DEG with a square lattice constant $a_{0}.$ We allow the spin texture
(if any) to have the bigger lattice constant $\sqrt{2}a_{0}\,$by considering
the order parameters $\left\langle \rho _{\alpha ,\beta }\left( \mathbf{G}%
\right) \right\rangle $ with reciprocal lattice vectors given by $\mathbf{G}=%
\frac{2\pi }{\sqrt{2}a_{0}}\left( n,m\right) ,$ where $n,m=0,\pm 1,\pm 2,...$
The \textit{density} unit cell has a lattice constant $a_{0}$ while the 
\textit{magnetic} unit cell has a lattice constant $\sqrt{2}a_{0}.$ For the
potential strength, we use $W_{0}=eV_{e}F\left( G_{1}\right) $ [see Eq. (\ref%
{motion})], where $G_{1}=2\pi /a_{0}$. The critical values of $W_{0}$ (not $%
V_{e}$) for the transition between the uniform and the modulated phases at
different value of $\Gamma $ are similar. Hereafter, we give all energies in
units of $e^{2}/\kappa \ell $. We make the important assumption that Landau
level mixing by both the Coulomb interaction and the external potential can
be neglected i.e. we work in the limit of a weak superlattice potential. We
also neglect disorder effect and work at zero temperature.

The ratio $\ell /a_{0}$ that enters the Hartree-Fock energy and equation of
motion for the single-particle Green's function is given by%
\begin{equation}
\frac{\ell }{a_{0}}=\sqrt{\frac{1}{2\pi }\frac{\varphi _{0}}{Ba_{0}^{2}}}%
\equiv \sqrt{\frac{\Gamma }{2\pi }},  \label{deux}
\end{equation}%
where $\varphi _{0}=hc/e$ is the flux quantum. The important parameter $%
\Gamma ^{-1}$ is the number of flux quanta piercing a \textit{density} unit
cell area. With this definition, the factor $F\left( G_{1}\right)
=e^{-G_{1}^{2}\ell ^{2}/4}=e^{-\pi \Gamma }.$ We limit our analysis to $%
\Gamma =q/p$ where $q$ and $p$ are integers with no common factors.

In Landau level $N=0,$ a Wigner crystal\cite{Wigner} with a triangular
lattice can form at sufficiently small filling factor $\nu $\cite{LamGirvin}%
. At $\nu =1,$ however, the ground state of the 2DEG is a uniform electron
liquid with full spin polarization i.e. a quantum Hall ferromagnet\cite{Moon}
(QHF) whose energy per electron is given by%
\begin{equation}
\frac{\left\langle H_{HF}\right\rangle }{N_{e}}=-\frac{\Delta _{Z}}{2}-\frac{%
1}{2}\sqrt{\frac{\pi }{2}}\left( \frac{e^{2}}{\kappa \ell }\right)
\label{fock}
\end{equation}%
(neglecting the kinetic energy that is a constant in $N=0$). The QHF remains
the ground state even when the Zeeman coupling goes to zero because a
perfect alignment of the spins minimizes the Coulomb exchange energy [the
second term on the right-hand side of Eq. (\ref{fock})]. In a uniform state,
the Coulomb Hartree energy is cancelled by the neutralizing uniform positive
background.

\subsection{Case $\Gamma \in \left[ 1/2,1\right] $}

We first consider the case $\Gamma \in \left[ 1/2,1\right] .$ Figure 1 shows
the ground state energy and spin polarization $S_{z}/\hslash $ as a function
of the potential $W_{0}$ for $\Gamma =1/2,2/3,3/4,4/5,1$ and for a Zeeman
coupling $\Delta _{Z}=0.015$. The ground state is spatially uniform and has
an energy $\left\langle H_{HF}\right\rangle $ and a spin polarization $S_{z}$
that remain constant until a critical field $W_{0}^{\left( c\right) }\approx
0.11.$ This uniform state is described by only one order parameter, i.e. $%
\left\langle \rho _{+,+}\left( 0\right) \right\rangle =1,$ and is fully spin
polarized i.e. the spin per electron is $S_{z}=\hslash /2.$ The
corresponding change in the density of states (DOS)\ with $W_{0}$ is shown
in Fig. 2. In the absence of external potential and Coulomb interaction, the
DOS has two peaks at energies $E_{\pm }=\pm \Delta _{Z}/2$ corresponding to
the two spin states. With Coulomb interaction, the Zeeman gap $\Delta _{Z}$
is strongly renormalized [see Fig. 2 (a)] as is well known. When the
external potential is present, the DOS\ for each spin orientation, has $p$
peaks corresponding to the number of subbands expected when an electron is
submitted to both a magnetic field and a weak periodic potential\cite%
{Hofstadter}. This is clearly visible in Fig. 2 (a),(d),(e) for $\Gamma
=1,2/3,4/5$. The external potential increases the width of the peaks in the
DOS and decreases the renormalized Zeeman gap (which is also the transport
gap). We remark that the rapid oscillations in some of the graphs at $\Gamma
=1$ are a numerical artefact. They depend strongly on the number of
reciprocal lattice vectors kept in the calculation.

\begin{figure}[tbph]
\includegraphics[scale=0.9]{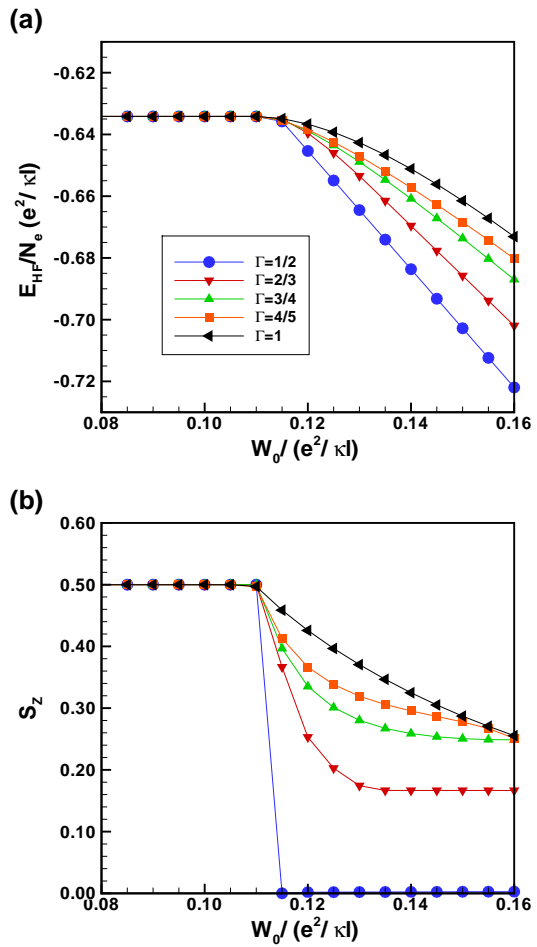}
\caption{(Color online) (a)\ Hartree-Fock energy per electron and (b)
average spin per electron $S_{z}/\hslash $ as a function of the applied
external field $W_{0}$ for different values of $\Gamma $ at filling factor $%
\protect\nu =1$ in Landau level $N=0$ and for the Zeeman coupling $\Delta
_{Z}/\left( e^{2}/\protect\kappa \ell \right) =0.015.$}
\end{figure}
%

If we enforce the uniform solution beyond the critical value $W_{0}^{\left(
c\right) }\approx 0.11$, we find that the transport gap closes at $%
W_{0}\approx 0.15$ for $\Gamma =1$ where the system becomes metallic. Our
code no longer converges in this case. But, this transition to a metallic
state does not occur because the uniform state becomes unstable at $%
W_{0}^{\left( c\right) }.$ The stability of a state is evaluated by
computing the dispersion relation of its collective modes. For the uniform
state, the collective excitations reduce to a spin-wave mode$.$ When $%
W_{0}=0,$ the spin-wave dispersion is given by\cite{Kallin}

\begin{figure}[tbph]
\includegraphics[scale=0.9]{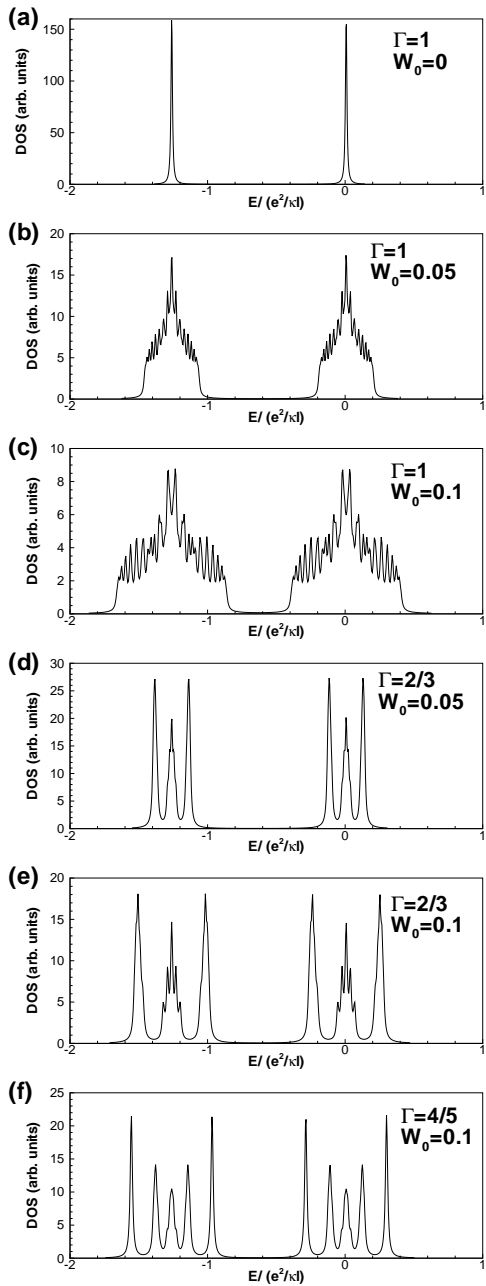}
\caption{Density of states for the uniform fully spin polarized phase at $%
\protect\nu =1$ in Landau level $N=0$ for Zeeman coupling $\Delta
_{Z}=0.015. $ (a)-(c) $\Gamma =1$ and $W_{0}=0,0.05,0.1$ respectively;
(d)-(e)~$\Gamma =2/3$ and $W_{0}=0.05,0.1$ respectively; (f) $\Gamma =4/5$
and $W_{0}=0.1$. The rapid oscillations in some of the graphs are a
numerical artefact. All energies are in units of $e^{2}/\protect\kappa \ell
. $}
\end{figure}

\begin{equation}
\omega _{SW}\left( \mathbf{k}\right) =\Delta _{Z}+\left( \frac{e^{2}}{\kappa
\ell }\right) \left[ X\left( 0\right) -X\left( \mathbf{k}\right) \right] .
\label{spinwave}
\end{equation}%
This mode is gapped at the bare Zeeman energy and saturates at $\omega
_{SW}\left( \mathbf{k\rightarrow \infty }\right) =\Delta _{Z}+\sqrt{\frac{%
\pi }{2}}\left( \frac{e^{2}}{\kappa \ell }\right) .$ Figure 3 shows its
dispersion for $\Gamma =1$ and $W_{0}=0,0.05,0.10,0.11.$ The wave vector $%
\mathbf{k}$ runs along the path $\Gamma -X^{\prime }-M^{\prime }-\Gamma $
i.e. along the edges of the irreducible \textit{density }Brillouin zone
(with $\Gamma =\left( 0,0\right) ;M^{\prime }=\left( 1/\sqrt{2},0\right)
,X^{\prime }=\left( 1/2\sqrt{2},1/2\sqrt{2}\right) $ in units of $2\pi
/a_{0} $). The spin-wave mode softens at a finite wave vector $\mathbf{k}$
as $W_{0} $ increases so that the uniform state becomes unstable at $%
W_{0}^{\left( c\right) }\approx 0.11$ which is also the value at which the
2DEG is seen to enter a new phase in Fig. 1(a). When plotted in the reduced
zone scheme as in Fig. 3, the spin-wave mode is split into several branches
that accumulate into a very dense manifold near $\omega _{SW}\left( \mathbf{%
k\rightarrow \infty }\right) $. Only some of these branches are shown in
Fig. 3 since we are interested only in the low-energy sector. The spin-wave
dispersion is obtained by following the pole of the response functions $\chi
_{\rho _{+},\rho _{-}}^{R}\left( \mathbf{k,k};\omega \right) $ with $\rho
_{\pm }=\rho _{x}\pm i\rho _{y}$ for different values of $\mathbf{k}.$ We
remark that the softening of the spin-wave mode by a \textit{one-dimensional}
external potential was reported previously by Bychkov et al.\cite{Bychkov}.
These authors suggested that the resulting condensation of the spin excitons
at the softening wave vector would create a new spin density wave ground
state. This is precisely what we find, but this time, for a \textit{%
two-dimensional} surface potential.

\begin{figure}[tbph]
\includegraphics[scale=0.9]{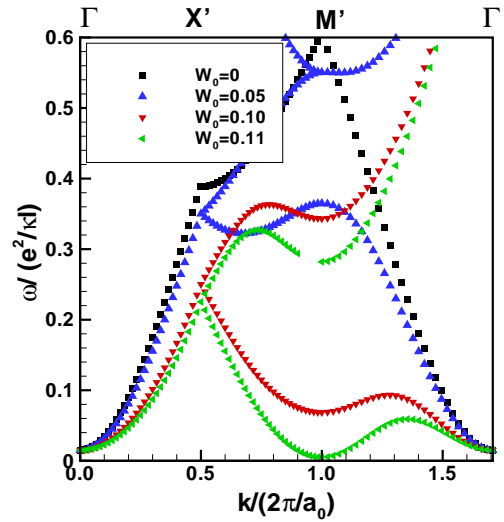}
\caption{(Color online) Dispersion relation of the spin-wave mode in the
uniform fully spin polarized state at $\protect\nu =1,N=0$ for $\Gamma =1,$ $%
\Delta _{Z}=0.015$ and different values of the external potentiel $W_{0}$.
The wave vector $\mathbf{k}$ follows the path: $\Gamma -X^{\prime
}-M^{\prime }-\Gamma $ along the irreducible (density) Brillouin zone of the
square lattice. All energies are in units of $e^{2}/\protect\kappa \ell .$}
\end{figure}

The ground state in a small region of $W_{0}$ after $W_{0}^{\left( c\right)
} $ is a charge density wave with a vortex spin texture. Hereafter, we refer
to this state as the vortex-CDW. The range of $W_{0}$ where the vortex-CDW
is the ground state depends on $\Gamma $ and $\Delta _{Z}.$The electronic
density and spin texture of the vortex-CDW are shown in Fig. 4 for the
parameters $\Gamma =1$ and $W_{0}=0.12,\Delta _{Z}=0.015.$ The value $%
W_{0}=0.12$ is close to $W_{0}^{\left( c\right) }$ so that the amplitude of
the CDW in this figure is small. The amplitude increases with $W_{0}$
however. Minima and maxima of the CDW have the same amplitude and there is
no net induced charge in a unit cell as expected. The spin density $%
S_{z}\left( \mathbf{r}\right) $ (not shown in the figure) varies only
slightly.

The spin texture of the vortex-CDW is interesting. There is a $2\pi $ spin
vortex at each positive and negative modulation of the density. Since the $z$
component of the spin is everywhere positive and because of the spin-charge
coupling inherent to a QHF\cite{Moon}, the positive and negative modulations
have opposite vorticity. We could, loosely speaking, refer to the positive
and negative modulations as merons and antimerons. A meron is an excitation
of a unit vector field $\mathbf{m}\left( \mathbf{r}\right) $ that has $%
m_{z}\left( 0\right) =\pm 1$ at its center and $m_{z}\left( \mathbf{r}%
\right) =0$ far away from the center where the vectors lie in the $xy-$plane
and form a vortex configuration with vorticity $n_{v}=\pm 1.$ As $r$
increases from the meron core, the spins smoothly rotate up (if $m_{z}\left(
0\right) =-1$) or down (if $m_{z}\left( 0\right) =+1$) towards the $xy-$%
plane. There are four flavors of meron\cite{Moon} with a topological charge
given by $Q=\frac{1}{2}\left[ m_{z}\left( \infty \right) -m_{z}\left(
0\right) \right] n_{v}.$ In a QHF, merons carry half an electron charge. In
our vortex-CDW however, we are dealing with a spin density $\mathbf{S}\left( 
\mathbf{r}\right) $ that does not have $\left\vert \mathbf{S}\left( \mathbf{r%
}\right) \right\vert =\hslash /2$ everywhere in space (the vectors do not
just rotate) so that our merons do not have a quantized charge. But, for a
given sign of $S_{z}\left( 0\right) $ the meron and antimeron have opposite
vorticity and so opposite electrical charge. Moreover, our merons are
closely packed in a square lattice and have a large core so that the spin
vector tilts towards the $xy-$plane but $S_{z}\left( 0\right) $ does not go
to zero between two adjacent merons.

In each magnetic unit cell of the vortex-CDW, there are two merons and two
antimerons with the same vorticity but opposite global phase for two merons
or antimerons. This bipartite meron lattice is similar to the square lattice
antiferromagnetic state (SLA) of the Skyrme crystal that was predicted to
occur in a 2DEG near (but not at) filling factor $\nu =1$ in the absence of
an external potential\cite{CoteSkyrme}. In the Skyrme crystal, the electrons
(or holes) added to the QHF state at $\nu =1$ crystallize in the form of
skyrmions for $\nu >1$ or antiskyrmions for $\nu <1$. The vortex-CDW that we
find here occurs \textit{at} precisely $\nu =1.$

Figure 1(b) shows that the average spin $S_{z}$ decreases with $W_{0}$ in
the vortex-CDW and saturates at the precise value $S_{z}=\Gamma -1/2$ for $%
\Gamma =1/2,2/3,3/4$. In the saturation region for $S_{z}$, the spin texture
has disappeared and the CDW has very little modulation in $S_{z}\left( 
\mathbf{r}\right) .$ We will call this phase, the normal-CDW. There is no
saturation for the two cases $\Gamma =4/5,1.$ We assume that this is due to
the fact that there is another phase very close in energy that wins over the
vortex-CDW for $W_{0}\gtrsim 0.16$ in these two cases (and probably at a
larger value of $W_{0}$ for the other cases) but we have not been able to
stabilize this other phase. We thus limit our analysis to the range $%
W_{0}\in \left[ 0,0.16\right] $ for most values of $\Gamma $ in this work.
The change in $S_{z}$ induced by the external potential should be detectable
experimentally. In particular, the vortex-CDW is absent for $\Gamma =1/2$
and thus the 2DEG makes a transition from a fully polarized to an
unpolarized CDW with period $a_{0}$ instead of $\sqrt{2}a_{0}$.

\begin{figure}[tbph]
\includegraphics[scale=0.9]{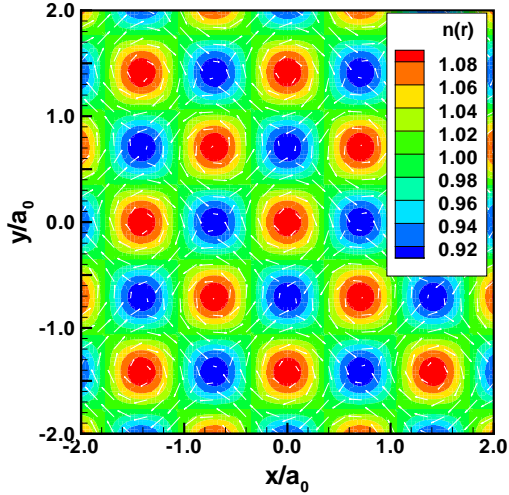}
\caption{(Color online) Vortex-CDW phase of the 2DEG at $\protect\nu =1$ in
Landau level $N=0$ for $W_{0}=0.12$ and Zeeman coupling $\Delta _{Z}=0.015.$
The electronic density $n\left( \mathbf{r}\right) $ is in units of $\left( 2%
\protect\pi \ell ^{2}\right) ^{-1}.$ The the vector field shows the vortex
structure in the $xy-$plane for the parallel component of the spin vector.
All energies are in units of $e^{2}/\protect\kappa \ell .$}
\end{figure}

The density of states for the vortex-CDW is shown in Fig. 5 for $\Gamma =2/3$
and $W_{0}=0.12,0.14,0.18.$ The subband structure gets more and more
different from that of the uniform phase as $W_{0}$ increases [compare with
Fig. 2(e)]. The electron-hole gap decreases slowly with $W_{0}$ in the
vortex-CDW and normal-CDW phases.

\begin{figure}[tbph]
\includegraphics[scale=0.9]{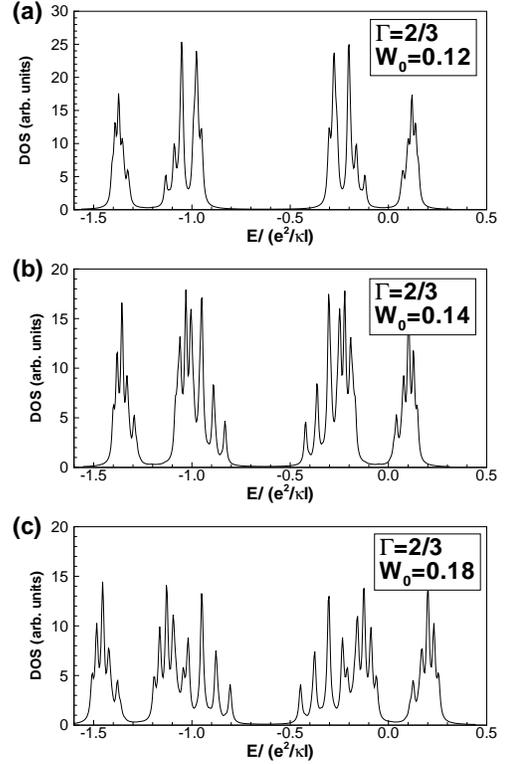}
\caption{Density of states in the vortex-CDW at $\protect\nu =1$ in Landau
level $N=0$ for $\Gamma =2/3$, Zeeman coupling $\Delta _{Z}=0.015$ and
different values of $W_{0}$ (all energies are in units of $e^{2}/\protect%
\kappa \ell $).}
\end{figure}

Figure 6 shows the Hartree, Fock (or exchange), external potential and
Zeeman contributions to the total energy of the vortex-CDW and uniform state
for $\Gamma =1$ and Zeeman coupling $\Delta _{Z}=0.015.$ The Hartree energy
is zero in the uniform phase and small in the vortex-CDW phase. The Zeeman
energy is also very small in both phases. The competition is between the
Fock and the external potential energies. The former is minimal in the
uniform state and increases with $W_{0}$ in the vortex-CDW phase. The latter
is zero in the uniform phase but decreases with $W_{0}$ in the vortex-CDW
state. Figure 6 shows that the increase in exchange energy is more than
compensated by the decrease in the external potential energy when the
vortex-CDW is formed.

\begin{figure}[tbph]
\includegraphics[scale=0.9]{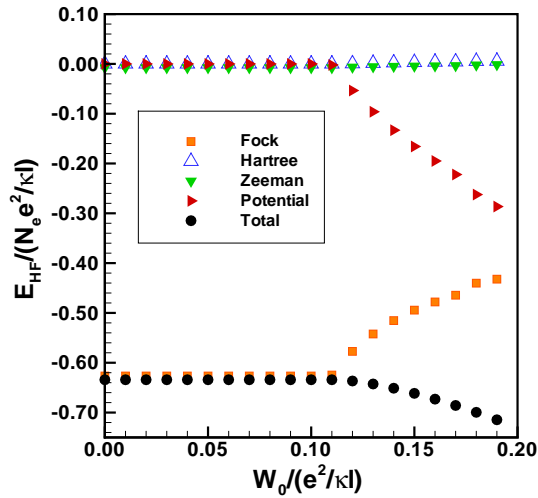}
\caption{(Color online) Behaviour with $W_{0}$ of different contributions to
the total energy of the uniform and vortex-CDW states for $\Gamma =1$ and
the Zeeman coupling $\Delta _{Z}=0.015.$}
\end{figure}

The energy of the vortex-CDW state does not depend on the global phase\ of
its vortices. This $U\left( 1\right) $ symmetry, which is broken in a
particular realization of the vortex-CDW state, leads to a gapless phase
mode (a Goldstone mode). This is clearly seen in Fig. 7(b) where the two
modes for $\omega <0.05e^{2}/\hslash \kappa \ell $ are the spin wave mode
which is gapped at $\Delta _{Z}$ and the gapless phase mode. In Fig. 7, $%
\Gamma =2/3$ and the wave vector $\mathbf{k}$ now follows the path $\Gamma
-M-X-\Gamma $ along the edges of the irreducible \textit{magnetic} Brillouin
zone of the square lattice [with $\Gamma =\left( 0,0\right) ;M=\left( 1/2%
\sqrt{2},1/2\sqrt{2}\right) ,X=\left( 1/2\sqrt{2},0\right) $ in units of $%
2\pi /a_{0}$]. To obtain the dispersions in the CDW phases, we have computed
the response functions $\sum_{\mathbf{G}}\chi _{\rho _{j},\rho
_{j}}^{R}\left( \mathbf{k+G},\mathbf{k+G},\omega \right) $ with $j=n,x,y,z$
keeping the first $25$ reciprocal lattice vectors in the summation. The
summation allows the capture of modes that originate from a folding of the
full dispersion into the first Brillouin zone. It also captures the
electron-hole continuum\cite{Cote2} that starts at the Hartree-Fock gap. In
Fig. 7, we have cut the dispersions at a frequency corresponding to the
onset of this continuum. Figure 7 shows the dispersion for: (a) the uniform
phase, (b) the vortex-CDW, and (c) the normal-CDW.

\begin{figure}[tbph]
\includegraphics[scale=0.9]{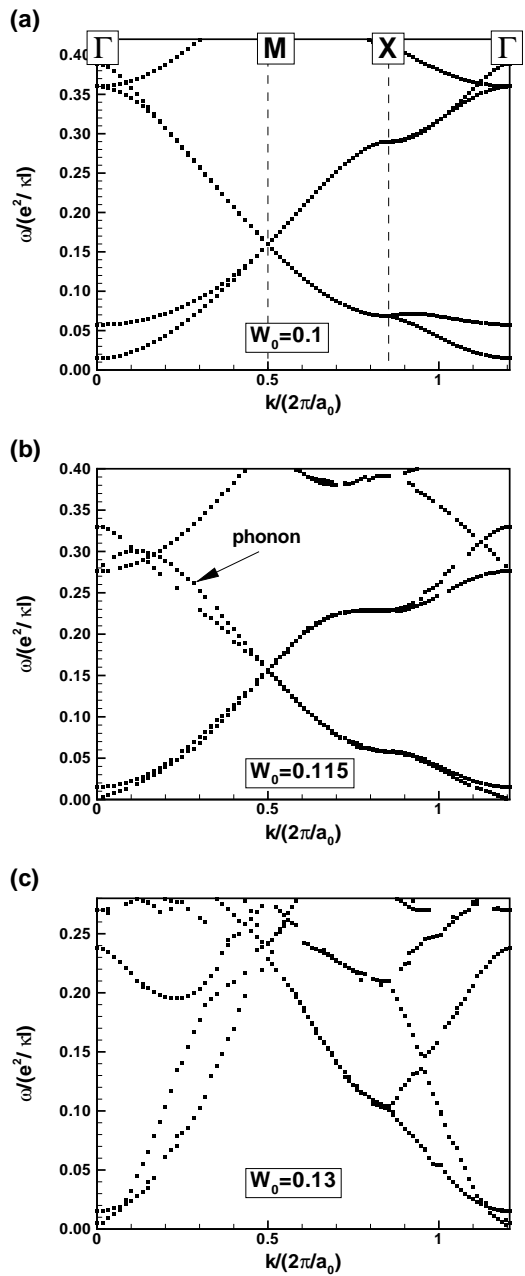}
\caption{Dispersion relations of the collective modes at $\protect\nu =1$ in
Landau level $N=0$ for $\Gamma =2/3$ and Zeeman coupling $\Delta _{Z}=0.015.$
(a) $W_{0}=0.1;$ (b) $W=0.115;$ and (c) $W=0.13.$ All energies are in units
of $e^{2}/\protect\kappa \ell .$}
\end{figure}

The vortex and normal CDWs have a phonon mode gapped by the external
potential. The branch we indicate as the gapped phonon mode in Fig. 7(b) has
the strongest peak in the response function $\chi _{\rho _{n},\rho
_{n}}^{R}\left( \mathbf{k},\mathbf{k},\omega \right) $ (no summation over $%
\mathbf{G}$) as $\mathbf{k}\rightarrow 0$ while the spin wave and phase
modes are stronger in $\chi _{\rho _{+},\rho _{-}}^{R}$ $\left( \mathbf{k},%
\mathbf{k},\omega \right) $ and $\chi _{\rho _{z},\rho _{z}}^{R}\left( 
\mathbf{k},\mathbf{k},\omega \right) $ respectively. At $W_{0}=0.13$ for $%
\Gamma =2/3,$ the ground state has transited to the normal-CDW and the phase
mode is gapped as shown in Fig. 7(c).

\subsection{Case $\Gamma \in \left[ 0,1/2\right] $}

The phase diagram for $\Gamma \in \left[ 0,1/2\right] $ is different from
that of $\Gamma \in \left[ 1/2,1\right] .$ For $\Gamma \in \left[ 0,1/2%
\right] ,$ we find a transition between two types of vortex-CDW phases. The
first vortex-CDW is the one described in the previous section, the second
one, the antivortex-CDW has the sign of all vortices and $S_{z}$ inverted
(but different amplitude for the charge and spin modulations). This
antivortex-CDW evolves from a uniform state that has all spin down as shown
in Fig. 8. At $\Delta _{Z}=0,$ these two CDW are degenerate in energy. At
finite Zeeman coupling, there is a crossing between the energy curves of
these two phases. The ground state thus evolves from the uniform state with
all spins up, to the vortex-CDW, then to the antivortex-CDW and finally into
the normal CDW. Figure 8 shows these transitions for the special cases of $%
\Gamma =1/3$ and $\Delta _{Z}=0.015,0.006,0.002.$ The corresponding behavior
of $S_{z}$ is also shown. The region where $S_{z}$ varies in each graph is
where the vortex-(or antivortex)-CDW is the ground state. As the Zeeman
coupling gets smaller, this region increases. The value of $W_{0}$ for the
crossing between the two vortex-CDW states is shown by the dashed vertical
line in the $S_{z}$ vs $W_{0}$ curves. The average spin $S_{z}$ changes
discontinuously at this point but this discontinuity goes to zero as $\Delta
_{Z}\rightarrow 0.$ The value of $S_{z}$ is always positive, however. For $%
\Gamma \in \left[ 1/2,1\right] ,$ the energy curve for the anti-vortex CDW
is above that of the vortex-CDW for all values of $W_{0}$. The two curves
merge at $\Delta _{Z}=0$ but there is no crossing between the two solutions.
If we take advantage of the possibility of changing the value of the $g-$%
factor independently of the magnetic field in GaAs/AlGaAs heterojunctions,
then it is possible to reduce the Zeeman coupling and, as Fig. 9 clearly
shows, to increase the transition region where the vortex-CDW is expected.

\begin{figure}[tbph]
\includegraphics[scale=0.9]{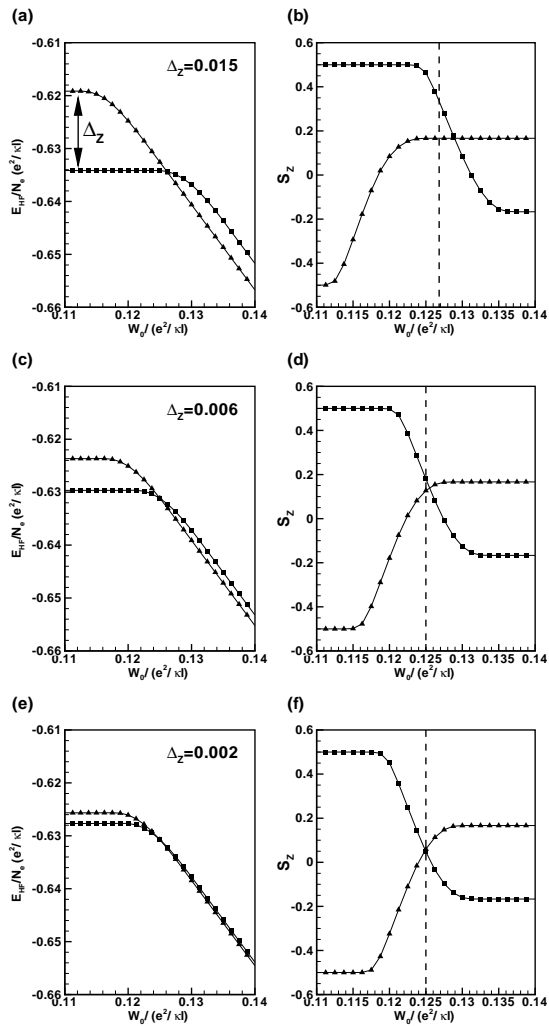}
\caption{Energy (left) and corresponding average spin $S_{z}/\hslash $ per
electron (right) of the vortex-CDW (squares) and antivortex-CDW (triangles)
for $\Gamma =1/3$ and Zeeman couplings : (a),(b) $\Delta _{Z}=0.015;$(c),(d) 
$\Delta _{Z}=0.006;$(e),(f) $\Delta _{Z}=0.002$ in units of $e^{2}/\protect%
\kappa \ell .$ The vertical dashed lines indicate the value of the potential 
$W_{0}$ at which the transition between the two vortex CDWs takes place.}
\end{figure}

Figure 9 shows the behavior of $S_{z}$ in the ground state for $\Gamma
=1/2,1/3,1/4,1/5$ and a very small Zeeman coupling $\Delta _{Z}=0.001.$ The
vertical dashed lines indicate where the transition between the vortex and
antivortex-CDW phases occurs for each value of $\Gamma .$ The spin starts at 
$S_{z}/\hslash =1/2$ in the uniform state then decreases in the vortex-CDW
(open symbols in Fig. 9). When the antivortex-CDW replaces the vortex-CDW as
the ground state of the system, the value of $S_{z}$ changes
discontinuously. This jump is more apparent for $\Gamma =1/5$ in Fig. 9.
After this discontinuity, $S_{z}$ increases (filled symbols in Fig. 9) until
it reaches the finite value $S_{z}/\hslash =\frac{1}{2}-\Gamma $ at large $%
W_{0},$ a value that is independent of the Zeeman coupling $\Delta _{Z}.$
The behaviour of $S_{z}$ is not monotonous. It the limit $\Delta
_{Z}\rightarrow 0,$ the $S_{z}$ curves for the vortex- and antivortex-CDWs
would cross at $S_{z}=0$ and there would be no discontinuity. In the special
case $\Gamma =1/2,$ the transition is directly from the uniform and fully
polarized state with $S_{z}/\hslash =1/2$ to the normal CDW where $S_{z}=0.$
There is thus an important discontinuity in $S_{z}$ in this case.

\begin{figure}[tbph]
\includegraphics[scale=0.9]{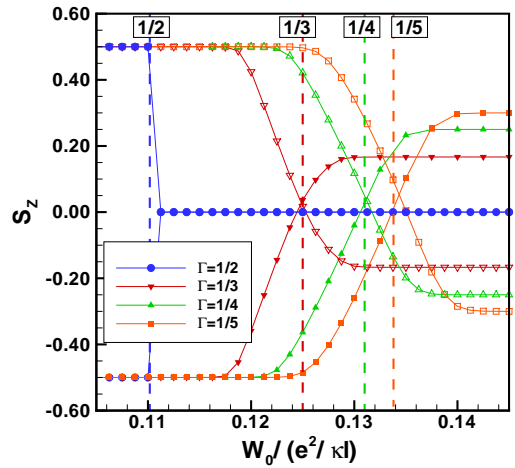}
\caption{(Color online) Spin polarization $S_{z}/\hslash $ as a function of
the applied external potential $W_{0}$ for several values of $\Gamma \leq
1/2 $ and Zeeman coupling $\Delta _{Z}=0.001e^{2}/\protect\kappa \ell .$ The
vertical dashed lines indicate the potential strength $W_{0}$ for each value
of $\Gamma $ where the transition from the vortex (filled symbols) to the
anti-vortex CDW (open symbols) takes place. For $\Gamma =1/2,$ the
transition is from the uniform (with $S_{z}/\hslash =1/2M)$ to the normal
CDW with $S_{z}=0.$}
\end{figure}

As we mentioned above, our formalism can equally well be used to discuss the
energy of the electron gas in Landau level $N=0$ in graphene if the
electrons are assumed to occupy only one valley$.$ An exact diagonalization
study by Ghazaryan and Chakraborty\cite{Tapash} for a 2DEG in graphene finds
transition between unpolarized and partially polarized ground states induced
by the external potential when $\Gamma =1$ (their $\alpha =1$). The equation
(\ref{motion}) was also used\cite{Wenchen} to study the effect of Coulomb
interaction on the density of states for graphene in a modulated potential
but the vortex-CDW state that we found was not considered in that work.

\section{SUMMARY AND DISCUSSION}

We have computed the phase diagram of the 2DEG at $\nu =1$ in Landau level $%
N=0$ in the presence of an applied external potential with a square lattice
periodicity for several rational values of $\Gamma \in $ $\left[ 0,1\right] $%
. We restricted our analysis to $W_{0}\in \left[ 0,0.16\right] .$ In this
range, the 2DEG evolves first from a uniform state with full spin
polarization then to a vortex-CDW and (if $\Gamma <1/2$) antivortex-CDW
state and finally into a normal CDW with no spin texture but with a finite
spin polarization $S_{z}$ if $\Gamma \neq 1/2.$ 

The change in the spin polarization $S_{z}$ with the applied field (smooth
for $\Gamma >1/2$ and abrupt for $\Gamma \leq 1/2$) is one feature of the
phase transition described in this work that should be measurable
experimentally. Another one is the gapless spin mode due to the broken U(1)
symmetry in the vortex-CDW phase. The same mode occurs in a Skyrme crystal.
In that system, it was shown that such mode could provide a fast channel for
the relaxation of the nuclear spin in nuclear magnetic resonance experiments%
\cite{Cote3}. Indeed, the bare Zeeman gap in the dispersion of the spin-wave
mode is orders of magnitude larger than the nuclear spin splitting, impeding
the creation of spin waves by nuclear spins. The softening of the spin wave
mode in the uniform phase may also lead to an increase in nuclear
spin-lattice relaxation time as suggested by Bychkov\cite{Bychkov}.

We have used $W_{0}$ for the external potential because the transition from
the uniform to the vortex-CDW takes place at roughly the same value of $W_{0}
$ when the potential is expressed in terms of $W_{0}$. The actual external
potential however is $V_{e}=F^{-1}\left( G_{1}\right) W_{0}=e^{\pi \Gamma
}W_{0}.$ This means that the critical field $W_{0}^{\left( c\right) }\approx
0.11$ translates into different real critical fields for different values of 
$\Gamma $ i.e. from $V_{e}^{\left( c\right) }=0.21$ for $\Gamma =1/5$ to $%
V_{e}^{\left( c\right) }=2.5$ for $\Gamma =1.$ It is not clear, then if our
assumption of neglecting Landau-level mixing can be justified for $\Gamma $
near unity. We assumed that the external lattice parameter $a_{0}$ is fixed
experimentally. When $\Gamma $ is also given, all other parameters are
determined: the magnetic field, the electronic density $n_{e}$ (at $\nu =1$)
and the ratio, $\alpha ,$ of the Coulomb interaction to the cyclotron
energy: 
\begin{eqnarray}
B &=&\frac{hc}{ea_{0}^{2}}\frac{1}{\Gamma }=\frac{4135.7}{\overline{a}%
_{0}^{2}}\frac{1}{\Gamma }\text{ T}, \\
n_{e} &=&\frac{1}{\Gamma a_{0}^{2}}=\frac{1}{\overline{a}_{0}^{2}}\frac{1}{%
\Gamma }\times 10^{14}\text{ cm}^{-2}, \\
\alpha  &=&\frac{\frac{e^{2}}{\kappa \ell }}{\hslash \omega _{c}^{\ast }}=%
\frac{a_{0}}{a_{B^{\ast }}}\sqrt{\frac{\Gamma }{2\pi }}=\frac{\overline{a}%
_{0}}{10.2}\sqrt{\frac{\Gamma }{2\pi }},
\end{eqnarray}%
where $a_{B}^{\ast }=\kappa \hslash ^{2}/m^{\ast }e^{2}$ is the effective
Bohr radius and $\overline{a}_{0}$ is the lateral superlattice constant in
nm. We used, for GaAs: $\kappa =12.9$ and $m^{\ast }=0.067m_{e}$ (where $%
m_{e}$ is the electron mass).

For $\Gamma =1,\alpha =0.039\overline{a}_{0}~$so that with a very small (but
physically feasible\cite{Melinte}) superlattice period of $a_{0}=39$ nm, we
get $\alpha =1.5,B=2.7$ T, $n_{e}=0.65\times 10^{11}$ cm$^{-2}$ while for $%
\Gamma =1/5,$ we get $\alpha =0.68,B=13.6$ T, $n_{e}=3.29\times 10^{11}$cm$%
^{-2}.$ The magnetic field and density pose no problem, but $\alpha $ is not
small, especially when $\Gamma >1/2.$ Clearly, a more sophisticated
calculation including a certain amount of Landau-level mixing and screening
is required to confirm that the vortex-CDW phase is effectively the ground
state in this system. We leave this to further work.

\begin{acknowledgments}
R. C. was supported by a grant from the Natural Sciences and Engineering
Research Council of Canada (NSERC). Computer time was provided by Calcul Qu%
\'{e}bec and Compute Canada.
\end{acknowledgments}

\end{document}